\pdfoutput=1

\documentclass[preprint]{aastex61}

\defcitealias{Boyajian16}{B16}		
\defcitealias{Neslusan17}{NB17}		
\defcitealias{Boyajian18}{B18}
\defcitealias{Wright16b}{WS16}

\shorttitle{Proper Motion of the Faint Star Near KIC~8462852}
\shortauthors{Clemens et al.}

\begin{document}

\title{Proper Motion of the Faint Star near \\
KIC~8462852 (Boyajian's Star) - Not a Binary System}

\author[0000-0002-9947-4956]{Dan P. Clemens}
\affiliation{Institute for Astrophysical Research, Boston University,
    725 Commonwealth Ave, Boston, MA 02215}

\author{Kush Maheshwari}
\altaffiliation{Boston University Research Internship in Science and Engineering (RISE) 2017 summer student}
\affiliation{Institute for Astrophysical Research, Boston University,
    725 Commonwealth Ave, Boston, MA 02215}
\affiliation{Wayzata High School, Plymouth, MN}

\author{Roshan Jagani}
\altaffiliation{Boston University Research Internship in Science and Engineering (RISE) 2017 summer student}
\affiliation{Institute for Astrophysical Research, Boston University,
    725 Commonwealth Ave, Boston, MA 02215}
\affiliation{North Hollywood High School, North Hollywood, CA}

\author[0000-0003-4609-0630]{J. Montgomery}
\affil{Institute for Astrophysical Research, Boston University,
    725 Commonwealth Ave, Boston, MA 02215}

\author[0000-0002-4958-1382]{A. M. El~Batal}
\affil{Institute for Astrophysical Research, Boston University,
    725 Commonwealth Ave, Boston, MA 02215}

\author[0000-0001-6584-9919]{T. G. Ellis}
\affil{Department of Physics \& Astronomy, Louisiana State University,
261-A Nicholson Hall, Tower Dr.
Baton Rouge, LA 70803-4001}

\author[0000-0001-6160-5888]{J. T. Wright}
\affil{Department of Astronomy and Astrophysics, 
Pennsylvania State University,
424 Davey Lab,
University Park, PA 16802}

\correspondingauthor{Dan P. Clemens}
\email{clemens@bu.edu}

\begin{abstract}
A faint star located 2~arcsec from KIC~8462852 was discovered in 
Keck~10~m adaptive optics imaging in the $JHK$ near-infrared 
(NIR) in 2014 by \citet{Boyajian16}. 
The closeness of the star to KIC~8462852 suggested 
the two could constitute a binary, which might have
implications for the cause of the brightness dips seen by {\it Kepler} 
\citep{Boyajian16} and in ground-based optical studies 
\citep{Boyajian18}. Here, NIR imaging
in 2017 using the Mimir instrument 
resolved the pair and enabled measuring their separation. 
The faint star had moved $67 \pm 7$~milliarcsec (mas) 
relative to KIC~8462852 since 2014. 
The relative proper motion of the faint star is $23.9 \pm 2.6$~mas~yr$^{-1}$,
for a tangential velocity of $45 \pm 5$~km~s$^{-1}$ if it is at the same
390~pc distance as KIC~8462852. Circular velocity at the
750~AU current projected separation is 
$1.5$~km~s$^{-1}$, hence the star pair
cannot be bound. 
\end{abstract}


\section{Introduction}

The F3V star KIC~8462852 (``Boyajian's Star''\deleted{hereafter KIC~846}) at a 
distance of 390~pc \citep{GAIA16} was found to be unusual
by citizen-scientists examining {\it Kepler} light curves for the Planet Hunters program 
\citep{Fischer12}. 
The star exhibited several strong brightness
dips as well as families of lesser dips, with dip durations longer than typical of
exoplanet transits \citep[][hereafter B16]{Boyajian16}.  
Many models have been offered 
(see reviews by \citetalias{Boyajian16} and \citet{Wright16b}), 
from alien megastructures \citep{Wright16a}, to swarms of nearly a thousand,
co-traveling comets \citep{Bodman16}, to dust-enshrouded massive objects 
on elliptical orbits \citep{Neslusan17}, to twin collections of dusty systems of 
Trojan asteroids leading and following a planet with a tilted ring system
\citep{Ballesteros18}. 

A fainter, possible companion, star (hereafter `FS') was discovered in Keck Adaptive Optics 
(AO) observations conducted on 2014 October 16 in the near-infrared (NIR) 
$J$ (1.25~$\mu$m), $H$ (1.64~$\mu$m),
and $K$-bands (2.20~$\mu$m) by \citetalias{Boyajian16}. 
Optical speckle observations of KIC~8462852 
did not detect FS, confirming its faint, red nature \citepalias{Boyajian16}.
FS appeared 4.2 ($J$) to 
3.6 ($K$) mag fainter than KIC~8462852, and was assigned a possible M2V classification
by \citetalias{Boyajian16}. They noted that the inferred large physical separation of FS
from KIC~8462852 ($\sim$~900~AU) meant the former star was unlikely to directly cause the deep brightness dips
of the latter, though either a slow passage of FS through the system or
a binary nature for the stellar pair could affect stability of bodies in the outer reaches
of the KIC~8462852 system. The proximity of FS to KIC~8462852 arising from
chance alignment of field stars was estimated by \citetalias{Boyajian16} to be $\sim$~1\%. 
The full nature of FS and its effects on the KIC~8462852 system remained unknown and
in need of additional NIR observations, as optical imaging had proved unable to
detect FS.

A 1.5-2\% dip event for KIC~8462852, of six days duration, began
on 2017 May 19 \citep{Boyajian17a}, triggering optical monitoring of the stellar 
brightness at many observatories \citep{Boyajian18}. Starting
shortly thereafter, on May 25, the Mimir multi-function,
NIR instrument \citep{Clemens07} was used to begin monitoring KIC~8462852
throughout the May/June/July period and on one night in November.
These observations examined the KIC~8462852 system for NIR $JHK$ photometric 
variability, $HK$-band 
low-resolution spectroscopic properties, as well as $H$ and $K$-band imaging 
polarimetry of KIC~8462852 and the other stars in its 
vicinity.

The NIR photometric variations, as well as the polarimetric and spectroscopic 
properties of KIC~8462852 resulting
from these observations, are the subjects of related work \citep{Clemens18}.
This paper uses the Mimir data to develop astrometric findings regarding this stellar pair.
Section~\ref{obs} presents a summary of the Mimir observations and the
data processing steps. Section~\ref{analysis} describes the
analysis of the image data to isolate FS and measure
its angular separation from KIC~8462852. Section~\ref{discussion}
compares the measured relative proper motion for FS to the tangential velocity it
would have in circular orbit about KIC~8462852 and assesses the impact of the
findings. Section~\ref{summary} recaps the study findings.

\section{Observations and Data Processing}\label{obs}

NIR $JHK$  observations were obtained using Mimir on the 
1.8~m Perkins telescope, 
located outside Flagstaff, AZ on multiple nights spanning UT 2017 May 25 though July 06. 
A second set of observations was obtained 
on 2017 November 16. Mimir employed a $1024 \times 1024$~ALADDIN~III InSb array detector, cooled
to 33.5~K, with reimaging optics cooled to 65-70~K. The plate scale was 0.58~arcsec per pixel,
resulting in a $10 \times 10$~arcmin field of view (FOV). Photometric imaging used 
the Mauna Kea Observatory NIR filter set \citep{Tokunaga05}. Polarimetry 
additionally utilized a rotatable, 
cooled, compound half-wave plate (HWP), for each of the $H$ and $K$-bands, to 
introduce polarization modulation and a fixed, cooled wire grid 
for analysis. All observations were auto-guided, scripted and under computer control, 
including telescope motions as well as filter and HWP orientation changes. 

Imaging photometry for the first observing run consisted of single 2.5~s exposures 
in each waveband, obtained toward six sky-dither positions, offset by 
15-20~arcsec. During May 25 through June 19, one 
set of $JHK$ observations was obtained per night. During later nights of that run, a mix
of one, two, or three 
observation sets were obtained each
night. Average \replaced{FWHM}{full-width at half-maximum} seeing values were 1.50, 1.47, and 1.33~arcsec in the $J$, $H$, 
and $K$ bands, respectively. For the November run, photometric imaging consisted 
of six sky-dithered exposures of 
10, 5, and 10 sec respectively in the $JHK$ bands. Fifteen observations 
were conducted in each waveband. Average seeing values were 
1.51, 1.37, and 1.30~arcsec in $JHK$.

Imaging polarimetry in the $H$-band (`$H$-pol') was performed on four nights in the May - July period
and was performed in the $K$-band on two nights in that period. These observations used 
5~sec (two May nights) or 10~sec (two June nights) integration times for each of 
sixteen HWP orientation angles at each of 
six sky-dither positions to comprise one observation. Multiple observations were obtained 
for two of the June nights. In-dome flat-fields obtained for each HWP orientation were 
used to calibrate, 
along with the same darks and linearity data as for photometry. 
The average seeing values were 1.36 and 1.58~arcsec
for $H$-pol and $K$-pol, respectively.

The processing steps for Mimir polarimetry data were described 
in \citet{Clemens12a} with calibration described in \citet{Clemens12b}. 
To summarize, the raw science data were transformed
into linearized data, dark and flat-field images were similarly 
linearized, and the darks and flat-fields were used to correct the transformed data into
science-ready images. These were grouped as 96-image observations to obtain astrometric image solutions using  the positions of 2MASS \citep{Skrutskie06} stars 
present in each image. Stars found in each image were matched across images to
obtain relative image shifts and photometric differences. 

Photometric processing skipped many of the polarization steps, as all images 
were obtained without any HWP or wire grid present in the optical beam. 
Each six-position sky-dither image in each 
waveband was analyzed to find detected stars and to measure their positions and
fluxes. These were matched to resolve sky transmission 
variations and to flag and reject images with poor seeing or high 
winds, prior to stacking and summing the images making up each observation.

Astrometric fitting of the stars found in the final summed images to the 
positions of 2MASS stars resulted in typical positional difference
standard deviations of about $60 - 80$~mas \citep{Clemens12a}. The plate scale and
field rotation angle values were established with uncertainties of one part in 2,000 or less,
contributing negligibly to the measured uncertainties in the relative angular separations 
and position angles of the two stars described in the following.

\section{Data Analysis and Findings}\label{analysis}

\begin{figure}
\includegraphics[trim=0in 0in 0 0in, clip, angle=0,scale=0.65]{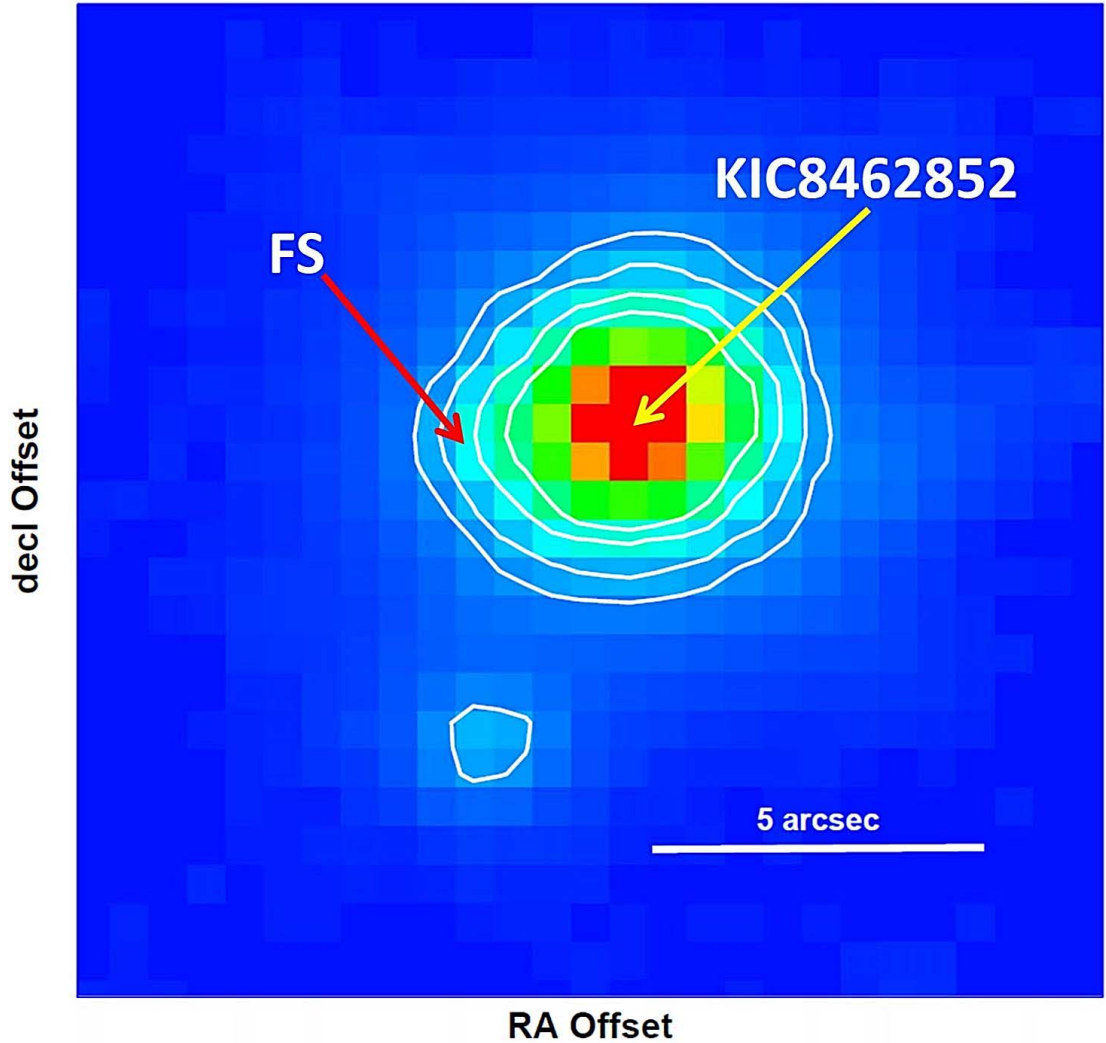}
\caption{\small
False color representation of a $15.5 \times 15$~arcsec portion of one of the 
Mimir $H$-band stacked observations, stretched to reveal the slight elongation to the
east caused by the FS star on the side of the PSF of KIC~8462852. The two stars are 
separated by about 2~arcsec and a 5~arcsec reference angle is at lower right. Individual
pixel sizes are 0.58~arcsec. White contours are stepped logarithmically to help show the
eastward elongation.
\label{fig_zoom}}
\end{figure}

FS, discovered two arcsec east of KIC~8462852 in the Keck AO observations of
\citetalias{Boyajian16}, appears as a distortion on the eastern side of the Mimir stellar 
profile of KIC~8462852, as seen in the zoomed portion of one of the $H$-band stacked
images in Figure~\ref{fig_zoom}. That figure shows the relative center locations of
FS and KIC~8462852, with a 5~arcsec reference indicated at lower right.  
Mimir data analysis tools include a \replaced{PSF}{point spread function (PSF)} modeling component that builds the master 
list of stars in the field using the multiple, blended star PSF fitting approach of DAOPHOT \citep{Stetson87}.
For sufficiently long exposures or averages of many such exposures, the star finding, PSF
modeling and stellar removal, and additional star finding processes enabled detection of 
FS in the Mimir images. 

The ninety-six images making up each 
$H$-pol and $K$-pol observation, when stacked and summed, were sufficiently 
deep to enable reliable separation of FS from KIC~8462852 for
twelve of the fourteen $H$-pol observations and for both $K$-pol observations. The 37 observations in each waveband for the short exposure photometric observations 
from the 2017 May/June/July observations yielded 29 detections of FS. The longer
exposure 2017 November photometric observations allowed separating FS from 
KIC~8462852 for 40 of the 45 observations. 

Collectively, the observations yielded 83 sets of equatorial
coordinates for FS and for KIC~8462852. These were differenced for every
observation to find relative RA and decl offsets. The offsets were grouped by
exposure time and waveband ($H$-pol and $K$-pol were the exceptions that
were grouped together) 
to yield seven data subsets for each of the RA and
decl offsets. The gaussian natures of the offset distributions 
were examined with a boot-strapped Kolmogorov-Smirnov approach.
This returned the likelihood that a data subset was not greatly different from a gaussian
characterized by the mean and dispersion of the data. Eleven of the fourteen
data sets had likelihoods exceeding 90\%, while the lowest likelihood was 40\%. 
The lower likelihood data subsamples showed somewhat more positional deviation
outliers than for a normal distribution and also tended to have
the shorter exposure times. The longer exposure data had likelihoods greater than
60\% and well-constrained positional scatter. Hence, positional uncertainties within each
data subset were set equal to the standard
deviation of the relative offsets for that data subset, separately for RA and decl.
The inverse variances of 
these uncertainties were used as weighting factors when forming offset averages and
propagated uncertainties. The offset standard deviations were
smallest for the polarimetry data ($\sim 7 - 15$~mas; longest  
integration times), moderate for
the longer exposure November photometry ($\sim 13 - 40$~mas), 
and largest for the short exposure photometry ($\sim 15 - 65$~mas).

\begin{figure}
\includegraphics[trim=0 0.5in 0 1.25in, clip, angle=0,scale=0.43]{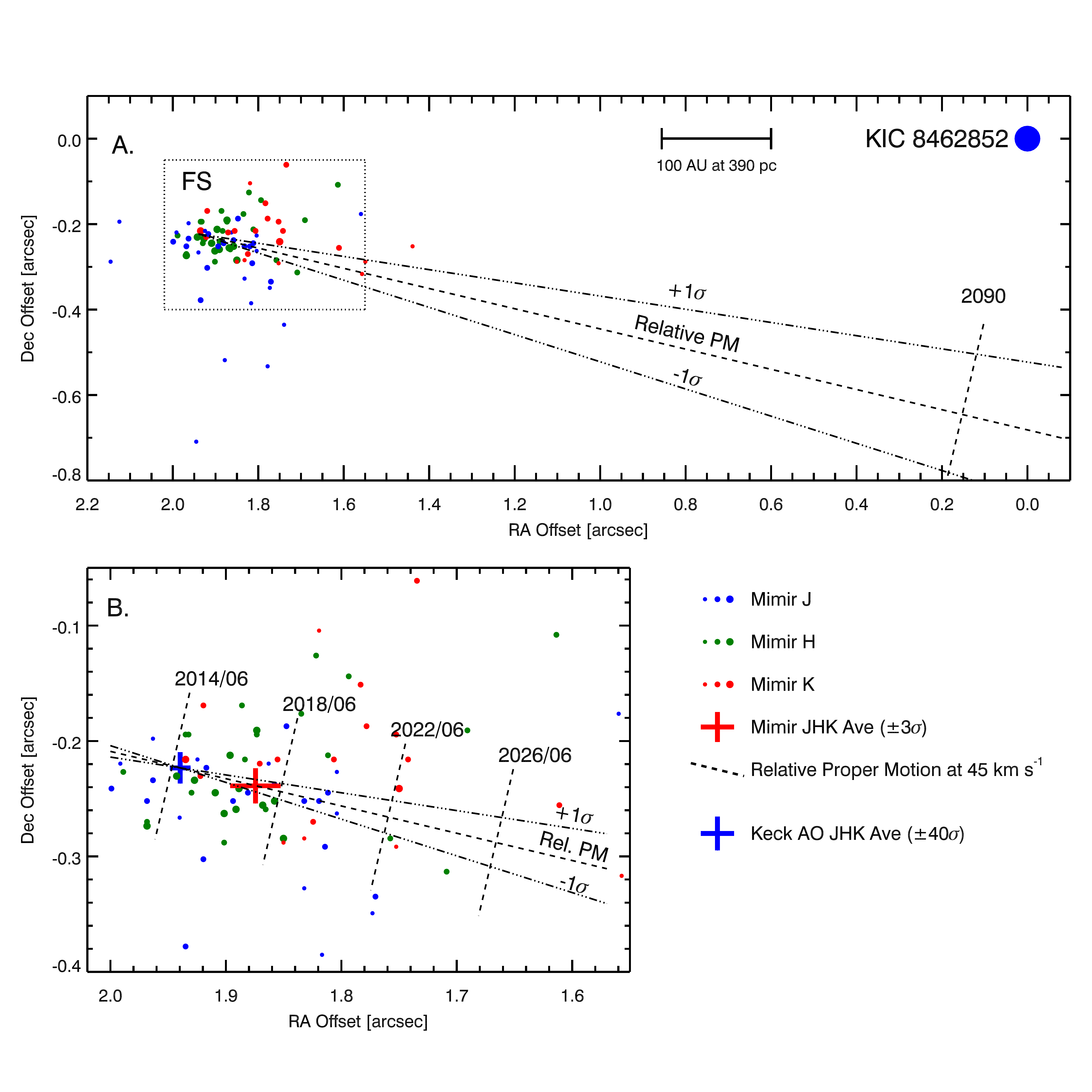}
\caption{\small
Sky offset positions for the faint star (FS) relative to KIC~8462852. Upper, A, panel
shows that the relative offsets for FS, measured from Mimir data, are projected to
lie some 750~AU southeast of KIC~8462852. 
A scale bar in the panel shows a length of 100~AU at 
390~pc distance. The dashed line is the
nominal relative proper motion vector. The dot-dashed lines show vectors
offset by the uncertainty in the vector orientation. Along the nominal vector, closest approach
to KIC~8462852 will happen in 2090, at a distance of about 260 AU. 
The lower, B, panel shows a zoomed
view of the dotted rectangular region in Panel A. Filled small, colored circles in 
both panels represent
measured Mimir relative offsets for FS. The average Mimir offset
location and associated
$3\sigma$ uncertainties are shown as the thick red cross. The average Keck AO $JHK$ position
\citepalias{Boyajian16} and forty times its uncertainties \citep{Boyajian17b} are shown
as the thick blue cross. The dashed line shows the 
relative motion vector for 45~km~s$^{-1}$, with perpendicular marks at 
four year intervals, on June centers. The proper motion measured corresponds to 
thirty times the circular velocity FS would have about KIC~8462852, if bound.
\label{fig_pm}}
\end{figure}

The relative positions measured for KIC~8462852 and FS are presented in 
Figure~\ref{fig_pm}. The upper, A panel
shows the Mimir-measured offsets, color-coded by waveband, with symbol
size indicating the relative weighting (the largest symbols represent 
data with the most weight).
The lower, B, panel shows
a zoom of the dotted rectangular region in Panel A. In Panel B, the Keck AO position 
reported by \citetalias{Boyajian16}, and updated with 
uncertainties from \citet{Boyajian17b}, is
shown as the thick blue error bars.
These are 40 times the uncertainties of the Keck AO positions
\citep{Boyajian17b}, to enable visualization. The weighted average offset and $3\sigma$ 
uncertainties for the Mimir observations are shown as the red cross. 
In addition to the nominal proper motion vector, vectors with position angles offset by
5\degr\ (1$\sigma$) are shown in both Figure panels. In the A panel, the date of closest
projected approach for the nominal relative motion vector is 2090, at which 
time FS will be about 260~AU from KIC~8462852, if both are at 390~pc distance.

A mean position shift of FS relative to KIC~8462852 with observing waveband was 
found, and can be seen in the
distributions of colored dots in Figure~\ref{fig_pm}. The blue, $J$ band dots have a 
tendency to be somewhat south and east of the $H$ band dots while 
the red, $K$ band dots continue
that trend to be found mostly to the north and west. A variance-weighted fit returns a 
waveband-position
vector with an equatorial position angle (EPA; measured east from north) 
of 307\degr\  of length 92~mas ($K$ to $J$), with a SNR of about 
three. This is likely due to some combination of the interactions among the red color of 
FS, the detailed PSF structure in each waveband, and the sampled seeing and focus. As there
is little astrophysical reason to believe a waveband-position gradient in the FS location should
be present, and the gradient found is not highly significant, the FS location could either
be reported as the fitted $H$ band value, say, or the weighted average of the positions
for all wavebands. These approaches return values identical to within a small fraction
of their uncertainties, so the averaging method was adopted for simplicity.

\begin{deluxetable}{lDD}
\tabletypesize{\scriptsize}
\tablecaption{KIC~8462852 to Faint Star Projected Offsets \label{tab_comp}}
\tablewidth{0pt}
\tablehead{
\colhead{Property}&\multicolumn2c{Boyajian (2017b)}&\multicolumn2c{This Work}\\
&\multicolumn2c{Keck AO}&\multicolumn2c{Mimir/Perkins}\\
&\multicolumn2c{$J$ $H$, $K$ Ave.}&\multicolumn2c{$J$, $H$, $K$ Ave.}\\ 
}
\decimals
\startdata
Mean Julian Date Offset & 0 & 1,024 \\ 
\ \ \ (relative to JD 2456947) \\ [6pt]
RA Offset [arcsec]&1.93957 &1.8743 \\
& (0.00023) & (0.0073) \\[6 pt]
decl Offset [arcsec]& $-0.22328$ & $-0.2388$ \\
& (0.00023) & (0.0051) \\[6 pt]
Radial Offset [arcsec] & 1.95237 & 1.8895 \\
& (0.00023) & (0.0115) \\[6 pt]
Equatorial Position Angle [deg]& 96.5667 & 97.26 \\
& (0.0067) & (0.16) \\ [10pt]
\hline
RA Offset Difference [mas] &\multicolumn4c{65.3 (7.3)}\\
decl Offset Difference [mas] &\multicolumn4c{15.5 (5.1)}\\
Offset Vector Amplitude [mas]&\multicolumn4c{67.1 (7.2)}\\
Proper Motion Amplitude [mas~yr$^{-1}$]&\multicolumn4c{23.9 (2.6)}\\
Proper Motion EPA [\degr\ E of N]&\multicolumn4c{256.7 (4.7)}\\
Tangential Speed (at 390~pc) [km~s$^{-1}$] &\multicolumn4c{44.9 (4.9)}\\
\enddata
\end{deluxetable}

The top portion of Table~\ref{tab_comp} presents the Mimir and Keck AO 
average relative offsets in both equatorial directions and as radial offsets and 
EPAs of FS from KIC~8462852. 
The bottom portion presents the differences in the KIC~8462852 to FS offset angles along the RA and decl directions between the Keck AO to Mimir observing
dates.
These yield an amplitude for the relative offset difference vector and the EPA of that vector. 
The effective date listed for the Mimir observations was formed from
the average of the dates weighted separately for the RA and decl offsets, using the same
weighting approach described above. 
Using this effective time separation between the Keck AO and 
Mimir observations, the relative proper motion amplitude was found to be about
24~mas~yr$^{-1}$. Under the assumption that both stars are at the 390~pc
distance, the tangential speed of FS, relative to KIC~8462852, was found to be 
$44.9 \pm 4.9$~km~s$^{-1}$.

In Table~\ref{tab_comp} and in Figure~\ref{fig_pm}, 
the Mimir RA uncertainties are larger than the decl 
uncertainties. This likely results from the uncertainty in the modeling and removal of the PSF of 
KIC~8462852, which overlaps FS significantly along the RA direction but less so along the decl direction. 

\section{Discussion}\label{discussion}

The proper motion of KIC~8462852 has been reported by \citet{GAIA16}, 
UCAC4 \citep{UCAC4}, and
Tycho-2 \citep{Tycho}. Weighted means of these reported values are 
$-11.9 \pm 0.5$~mas~yr$^{-1}$
along RA and $-10.2 \pm 0.9$~mas~yr$^{-1}$ along decl.
These constitute a projected proper motion of $15.7 \pm 0.7$~mas~yr$^{-1}$ along
EPA $229.\degr4 \pm 2.\degr5$ at a tangential velocity of 
$29.4 \pm 1.3$~km~s$^{-1}$.
These are similar to, but distinct from, the relative proper motion between FS
and KIC~8462852 (45~km~s$^{-1}$ along EPA $\sim$260\degr). The projected vector sum
results in absolute proper motions for FS of $-35.3 \pm 2.6$~mas~yr$^{-1}$
along RA and $-15.7 \pm 2.0$~mas~yr$^{-1}$ along decl, or $72 \pm 5$~km~s$^{-1}$
directed along EPA 246\degr. 

The distance to FS may be resolved by GAIA observations, but that
depends on its spectral type. If it is the M2V value suggested
by \citetalias{Boyajian16}, it should appear as a $V \sim 18$ mag object next to the
11.5~mag KIC~8462852. If, instead, the spectral type is M5.5V or later, its apparent
optical magnitude may fall below the sensitivity limit of GAIA. 

\citetalias{Boyajian16} also estimated the duration of passage of FS 
through the KIC~8462852 system, assuming identical distances and a relative motion of
10~km~s$^{-1}$, to be of order 400 years. Here, the updated tangential speed is
4-5 times greater, yielding a much shorter passage. This should
reduce the likelihood for scattering of objects from their outer KIC~8462852 system orbits
that might have led to star-grazing or star-plunging comets or other bodies 
\citep{Bailey92,LDE99a,LDE99b,Bodman16}.

The maximum circular orbit speed for FS,
at the apparent projected 750~AU offset from KIC~8462852, would be 1.5~km~s$^{-1}$. 
The Keck-to-Mimir tangential velocity for the pair at a common distance of 390~pc 
exceeds this value by a factor of thirty. The stellar pair cannot be bound and so do 
not constitute a binary.

\section{Summary}\label{summary}

The relative proper motion of the faint star (FS) found within two arcsec of KIC~8462852
was measured relative to that brighter, and enigmatic, star by combining 
published Keck AO near-infrared imaging from 2014 with new Mimir near-infrared
imaging from 2017. The relative separations of 
FS from KIC~8462852 were measured in 83 Mimir observation sets and combined
to find that FS had moved by nearly 70~milli-arcsec over the three year interval.
If FS is at the 390~pc distance to KIC~8462852, then
the implied tangential velocity of 45~km~s$^{-1}$ and projected vector direction will
put closest projected separation from KIC~8462852 at 260~AU about 70 years from now.
The circular velocity 
at the projected 750~AU separation is only 1.5~km~s$^{-1}$, which is much less
than the implied tangential speed measured, if both stars are at the same distance. 
If the spectral
type is significantly later than M2V, then FS is less distant and would be unrelated to KIC~8462852.
In either scenario, the two stars are not in a bound pair, so models invoking KIC~8462852
brightness changes from interactions with FS are weakened.

\acknowledgments

The authors thank T. Boyajian and the anonymous reviewer for their comments
and suggestions. 
The Boston University Research Internships in Science and Engineering (RISE) program 
for accomplished rising high school seniors selected and supported the participation 
of K.M. and R.J. on this project. 
This publication made use of data products from the Two Micron All Sky Survey, 
a joint project of the University of Massachusetts and the Infrared 
Processing and Analysis Center/California Institute of Technology, funded by 
NASA and NSF. 
This research was conducted in part 
using the Mimir instrument, jointly developed at Boston University and Lowell 
Observatory and supported by NASA, NSF, and the W.M. Keck Foundation.
Support for the Mimir instrument and this scientific 
effort have been made possible by grants 
AST 06-07500, AST 09-07790, AST 14-12269 from NSF/MPS 
and NNX15AE51G from NASA to 
Boston University and by Perkins telescope
observing time awarded as part of the Boston University -- Lowell Observatory 
Discovery Channel Telescope partnership.

\facility{Perkins}

\end{document}